\documentclass[epj]{webofc}
\usepackage[varg]{txfonts}   
\usepackage{comment,color}

\def\be{\begin{equation}}
\def\ee{\end{equation}}
\def\bdm{\begin{displaymath}}
\def\edm{\end{displaymath}}

\woctitle{MESON2018 - the 15$^\textrm{th}$ International Workshop on Meson Physics}
\begin{document}
\selectlanguage{english}
\title{The $Y(4260)$ and $Y(4360)$ enhancements\\
 within coupled-channels}

\author{Susana Coito\inst{1}\fnsep\thanks{\email{scoito@ujk.edu.pl}}
}

\institute{Jan Kochanowski University, 25-406 Kielce, Poland
          }

\abstract{
Puzzling structures have been observed in the charmonium energy region, namely the $Y(4260)$ and the $Y(4360)$, that cannot be easily accommodated within quark model frameworks. The proximity of nearby dominant hadronic thresholds suggests that they play an important role in the formation of the enhancements. We present results of an unitarized effective Lagrangian model, where mesonic loops, equivalent to coupled-channels, and charmonium vectors $\psi$ interplay to generate line-shapes and poles.
}

\maketitle

The number of vectorial $Y$ enhancements seen in the experiment is more than the $\psi$ states that can be accommodated within the quark model. Recently, some of such $Y$ signals have been seen in non-dominant hadronic decay channels, while they have not been observed in the dominant ones, in spite of their proximity to some of them \cite{pdg}. The fact that the width of the $Y$'s is large is therefore a puzzle within the dominantly charmonium picture, since the suppressed decays alone, together with the weak and electromagnetic decays, could not account for such a short lifetime of these ``states''. Furthermore, there seems to be an identification problem of the resonances, since the peaks appearing in different channels are not exactly coincident. In some cases, proper analysis of the interferences with the background are even missing, and mere fitting with Breit-Wigner line-shapes is employed in a non reliable way, leading to the announcement of even more $Y$ peaks than what probably there actually are. Indeed, concerning the position of the known $\psi$ states, one can easily see, for instance in some $D\bar{D}\equiv DD$ data from Belle in Ref.~\cite{prd77p011103}, that the position of the $\psi$ states does not correspond necessarily to the position of the peaks (henceforth we omit the ``bar'' sign for antiparticles). Therefore, a much more reliable analysis of the $\psi$ masses has been established by fitting the invariant mass distribution to all hadrons over the invariant mass distribution to leptons, i.e., the $R$ parameter \cite{plb660p315}. The same sort of analysis would be recommended to the $Y$ states.

The possibility that some of the $Y$ enhancements might be companion poles of the dominant $\psi$ states is worth further studies. In the light sector, mechanisms of dynamical generation of states have been studied thoroughly, as it can be found in Refs.~\cite{prd65p114010} and \cite{zpc30p615,ijtpgtno11p179} or, within the approach we present here, in \cite{prd93p014002,npb909p418}. The same phenomenon has been studied for open-charm scalar mesons in Ref.~\cite{prl91p012003}, where the $D_0^*(2100-2300)$ has been predicted, and for scalar charmonia in Refs.~\cite{prd76p074016,epja36p189}, where the authors have predicted a new resonance at 3.7 GeV. 

The $Y(4260)$ was first detected in BaBar \cite{prl95p142001} in the decay $e^+e^-\to J/\psi\pi^+\pi^-$, and its estimated mass and width is now $M\sim 4.23$ GeV and $\Gamma=55\pm 19$ MeV \cite{pdg}. The proximity of the $Y(4260)$ to the $D_s^*D_s^*$ threshold, together with the fact that the signal has not been seen in any of the dominant open-charm decay channels, suggests that, rather than a state on its own existence, the $Y(4260)$ results from some interference among the $D_s^*D_s^*$ and other hadrons \cite{prl105p102001,prd79p111501}. The possible dynamical generation of the $Y(4260)$ has been studied in Ref.~\cite{prd80p094012}, where the Faddeev equations have been solved. A peak was found at about 4.15 GeV, which is rather closer to the $\psi(4160)$. An idea in which the $Y(4260)$ would be a ``molecular'' state, with a $c\bar{c}$ $D$-wave core coupled to channel $DD_1$, may be found in Ref.~\cite{prd96p114022}. Moreover, it is possible that the $Y(4360)$, on its turn, might be generated by a similar mechanism as the $Y(4260)$, where the dominant interference would involve the $DD_1+DD'_1$ thresholds.

Particularly interesting is the fact that several $\psi$ states do not show up in $J\psi\pi^+\pi^-$ and $h_c\pi^+\pi^-$ cross section data, as it can be verified in Refs.~\cite{prl118p092001,prl118p092002}, indicating interference patterns, as it was pointed out in Ref.~\cite{prl105p102001}. In fact, a simple interference analysis between the $\psi(4160)$ and $\psi(4415)$ states, using Breit-Wigner line-shapes, shows how both these $\psi$'s are vanished from the spectrum, giving place to a single distorted $Y(4220)$, that could actually be identified with the $Y(4260)$ \cite{epjc78p136}. Here, we present a preliminary result where the cross section to channel $J/\psi f_0(980)$ is computed using an effective Lagrangian approach, with open-charm meson-meson loops, and the $\psi(4160)$ as the dominant propagator. By noticing from \cite{pdg}, that the actual mass difference between the $\psi(4160)$ and $Y(4260)$ is only about 30 MeV, rather than 100 MeV, we check if a simple loop effect could shift the $\psi(4160)$ peak to a higher mass in channel $J/\psi f_0(980)$, case in which the same pole would be responsible for both the $\psi(4160)$ and $Y(4260)$ line-shapes. In addition, we present previous results of the same model for the $\psi(3770)$ and the $\psi(4040)$, where dynamically generated poles have been found together with the corresponding seed poles, originating distortions on the line-shapes.\\ 

The effective Lagragian model we employ here considers the process $e^+e^-\to\gamma\to\psi\to m_1m_2$, where $m_i$ are the final mesons. Crucial within this model is the definition of the propagator $\Delta(s)$, that includes the convergent sum over each meson-meson one-loop, i.e.~$(m_1m_2)_j$ channel, and a sum over $N$ different channels. The loop function $\Pi(s)$, where $s$ is the invariant energy squared, is then given by
\be
\Pi(s)=\sum_j^N \Big( \Omega_j(s) + i\sqrt{s}\Gamma_j(s)\Big),\ \ \Omega,\ \Gamma \in \Re\ ,
\ee
where the real part is given by the dispersion relations
\be
\Omega_j(s,m_1,m_2)=\frac{PP}{\pi}\int_{s_{thj}}^{\infty}\frac{\sqrt{s'}\Gamma_j(s^{\prime},m_1,m_2)}%
{s^{\prime}-s}\ \mathrm{d}s^{\prime},
\ee
and the imaginary part is given by 
\be
\Gamma_{\psi\to(m_1m_2)_j}(s)=\frac{k_j(s,m_1,m_2)}{8\pi s}
|\mathcal{M}_{\psi\to(m_1m_2)_j}|^{2}\ ,
\ee
\be
\label{amp}
|\mathcal{M}_{\psi\to(m_1m_2)_j}|^{2}=\mathcal{V}_j(s,m_1,m_2)f_{\Lambda}^{2}(q_j^2),
\ee
where $k_j$ is the final state momentum, $\mathcal{V}$ is the vertex amplitude computed using the Feynmann rules, and $f$ is a vertex form factor that depends on a cutoff parameter $\Lambda$, and on the off-shell momentum $q_j$, and it is here used as an exponential. The propagator comes as
\be
\label{prop}
\Delta(s)=\frac{1}{s-m_{\psi}^{2}+\Pi(s)},
\ee
and the unitarized spectral function is given in function of the energy $E$ by 
\be
d_{\psi}(E) =-\frac{2E}{\pi}\mathrm{Im}\ \Delta(E).
\ee

\begin{figure}
\begin{center}
\resizebox{!}{160pt}{\includegraphics{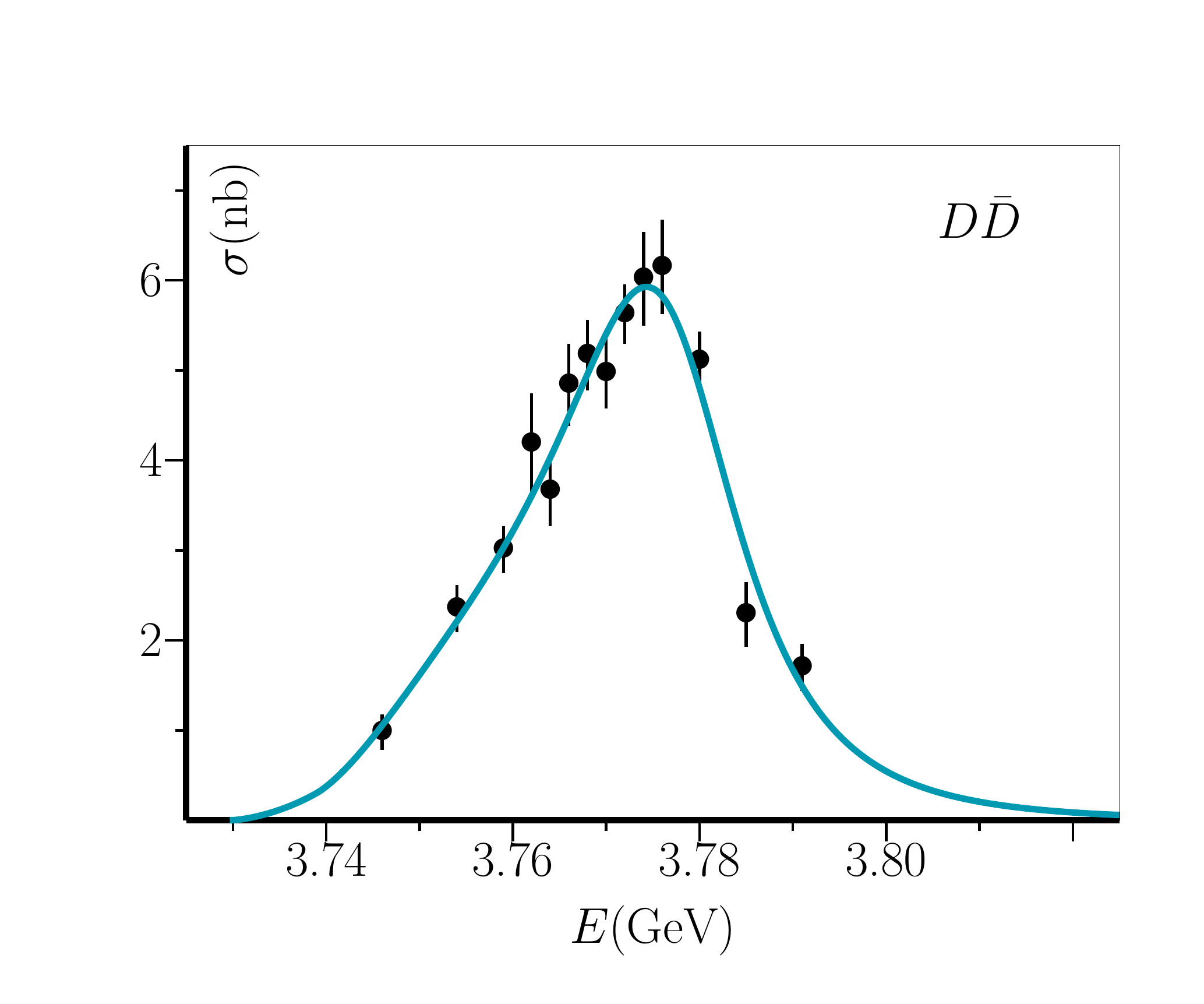}}
\caption{\label{fig1} Fit result of the unitarized effective model in \cite{1712.00969} to data in \cite{plb668p263}.}
\end{center}
\end{figure}

In Fig.~\ref{fig1} we show our fit to cross section data for the $\psi(3770)$, where we have used channels $D^+D^-$ and $D^0\bar{D}^0$ in the loops. Four free parameters have been used: the cutoff $\Lambda$, the seed mass $m_\psi$ entering in Eq.~\eqref{prop}, an effective coupling for the vertices $\psi DD$ (which is the same for both channels), and an effective coupling for the vertex $\psi e^+e^-$ that encloses the annihilation-creation process via photon. Beneath the $\psi(3770)$ structure we have found two poles, at $3741-i18$ MeV and at $3777-i12$ MeV, the first one generated dynamically, and the second one coming from the seed. These results may be found with great detail in Ref.~\cite{1712.00969}.

Concerning the $Y(4040)$, a new result, within the same model, predicts a dynamically generated resonance at about 4.00 GeV \cite{piotrowska}.

\begin{figure}
\begin{center}
\resizebox{!}{160pt}{\includegraphics{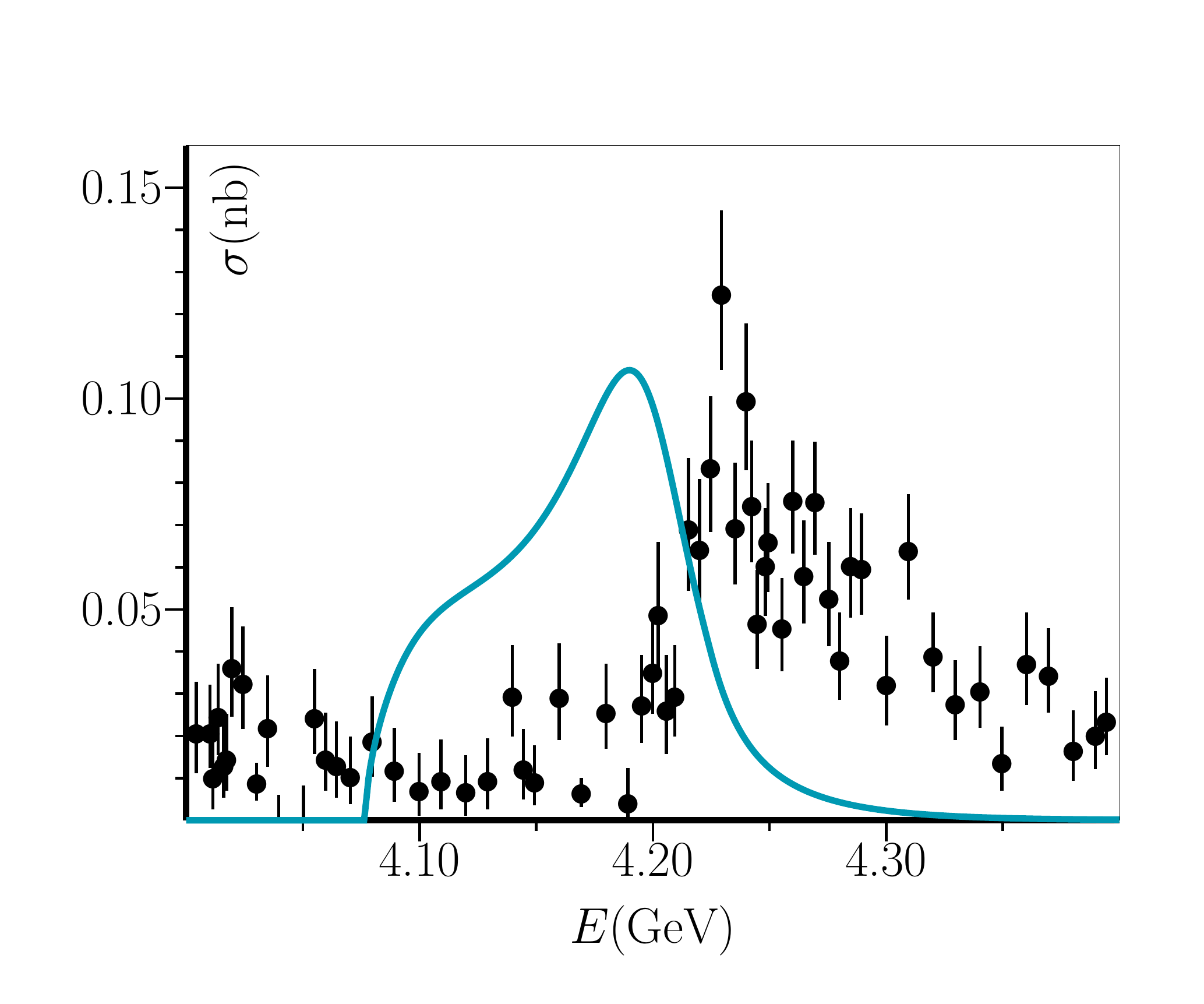}}
\caption{\label{fig2}Channel $J/\psi f_0(980)$ compared to $J/\psi \pi^+\pi^-$ data in Ref.~\cite{prl118p092001}.}
\end{center}
\end{figure}

Finally, we perform a calculation using channels $DD$, $DD^*$, $D^*D^*$, $D_sD_s$, $D_sD_s^*$ $D_s^*D_s^*$, and $J/\psi f_0(980)$. The couplings for vertices $\psi (m_1m_2)_j$ are computed using the partial decay ratios in \cite{pdg}, so as to reproduce a peak with mass and width corresponding to the $\psi(4160)$. The remaining free parameters are the cutoff, the seed mass for the $\psi$, and an amplitude factor for the cross section. A preliminary result is shown in Fig.~\ref{fig2}, where we clearly see that the dynamical effect of the intermediate loops distorts the line-shape, but does not shift the position of the $\psi(4160)$ peak. From here we conclude that either the $\psi(4160)$ and the $Y(4260)$ could still be the same resonance, but explained by some further effect, or they are in fact two structures with different origin. Concerning the $Y(4360)$, additional channels should be included, namely the $DD_1+DD_1'$, that falls over the peak. Other difficulty concerns the computation of the partial couplings to such higher energy channels. A systematic partial coupling scheme would be desirable to endue our current effective model.\\

In conclusion, we have presented an unitarized model that includes meson-meson loops and one seed, that is able to generate extra poles from the continuum, i.e., dynamical poles, which consequently distort the line-shapes, e.g., the case of the $\psi(3770)$. However, in order to account for the possible shifting of a peak position in different channels, some further interference mechanism needs to be contemplated.\\

The author thanks to F.~Giacosa for useful discussions. This work was supported by the
\textit{Polish National Science Center} through the project OPUS no.~2015/17/B/ST2/01625.




\end{document}